\documentclass[reprint,amsmath,amssymb,aps,prb]{revtex4-1}
\usepackage{natbib}
\usepackage{makeidx}
\usepackage{url}
\usepackage[colorlinks=true,linkcolor=blue]{hyperref}

\usepackage{graphicx}
\usepackage{bm}

\begin{document}

\preprint{APS/123-QED}

\title{Temperature induced collapse of spin dimensionality in magnetic metamaterials}

\author{Bj\"orn Erik Skovdal}
\email{bjorn\_erik.skovdal@physics.uu.se}
\author{Nanny Strandqvist}
\author{Henry Stopfel}
\author{Merlin Pohlit}
\author{Tobias Warnatz}
\author{Samuel D. Sl\"oetjes}
\author{Vassilios Kapaklis}
\author{Bj\"{o}rgvin Hj\"{o}rvarsson}

\affiliation{Department of Physics and Astronomy, Uppsala University, Box 516, 751 20 Uppsala, Sweden}

\date{\today}

\begin{abstract}
Spin and spatial dimensionalities are universal concepts, essential for describing both phase transitions and dynamics in magnetic materials. 
Lately, these ideas have been adopted to describe magnetic properties of metamaterials,  
as well as to replicate and explore ensembles of mesospins belonging to different universality classes. 
Here, we take the next step by investigating magnetic metamaterials not conforming to the conventional framework of continuous phase transitions. Instead of a continuous decrease in the moment with temperature, discrete steps are possible, resulting in a binary transition in the interactions of the elements.  
The transition is enabled by nucleation and annihilation of vortex cores, shifting topological charges between the interior and the edges of the elements. 
Consequently, the mesospins can be viewed as shifting their spin dimensionality, from 2 (XY-like) to 0, (vortices) at the transition.
The results provide insight into how dynamics at different length-scales couple, which can lead to thermally driven topological transitions in magnetic metamaterials.
\end{abstract}

\maketitle

\section{INTRODUCTION}
\hspace{\parindent}
Magnetic metamaterials \cite{Heyderman_Review_2013} composed of mesospins as building blocks \cite{ostman_ising-like_2018}, offer the possibility of tailoring magnetic interactions and dynamics in almost arbitrary ways. 
Previous investigations have mainly been focused on the collective magnetic order, dynamics and more exotic aspects such as frustration of artificial spin systems \cite{Heyderman_Review_2013, Nisoli_2013, Nisoli:2017hg, Rougemaille_2019}, while the internal magnetisation and dynamics of the mesospins are less explored \cite{Gliga_PRL_2013, Gliga_PRB_2015, Sloetjes_arXiv_2020}. For instance, extensive efforts have been made to mimic magnetic systems of various spatial and spin dimensionalities, where the shape of the elements have been used to enforce mesospins to be Ising- or XY-like \cite{ostman_ising-like_2018,sendetskyi_continuous_2019,ewerlin_magnetic_2013,streubel_spatial_2018,leo_collective_2018,arnalds2016new}. Even systems consisting of \textit{both} Ising- and XY-mesospins have been fabricated and investigated, which is a testament to the versatility of metamaterials  \cite{Arnalds_XY,ostman_interaction_2018}. The common denominator in all these investigations, however, is that the spin dimensionality is treated as a static property of the elements. Yet, the mesospins offer access to continuous degrees of freedom and rich internal magnetic states that go well beyond their atomic analogues. This attribute is the focal point of the present study, as we turn our attention to arrays of mesospins that can exhibit two distinct magnetisation states: collinear and vortex spin textures \cite{shinjo_magnetic_2000, Klaui_vortx_2003}. 
The mesospins can be thought of as having variable spin dimensionality, where the interaction strength depends on the inner magnetic textures. It is therefore possible to couple the changes in spin dimensionality to changes in the collective properties of the mesospins. The coupling of these internal and external degrees of freedom, in combination with the exploration of the role of topological effects on the observed transitions \cite{Mermin:1979io, Tchernyshyov:2005gs}, is the main motivation behind the current work.

\section{MATERIALS AND METHODS}

\subsection{Sample manufacturing}
\begin{figure}[t!]
\begin{center}
\includegraphics[width=1\linewidth]{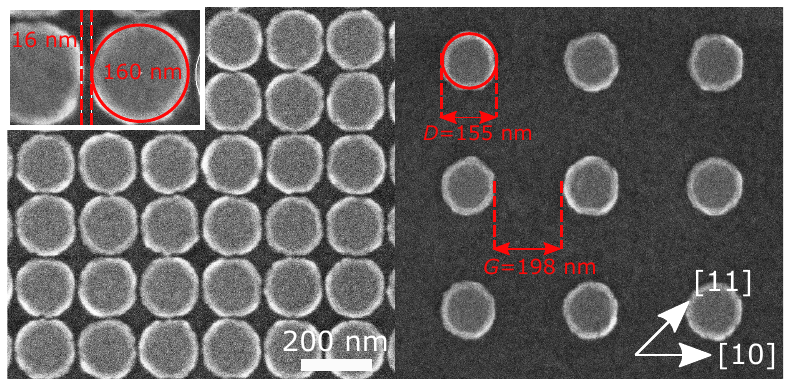}
\caption{The magnetic metamaterial - square arrays of circular islands. Scanning electron microscopy image of a representative structure with disks of a diameter $D = 150$ nm. (Left) Interacting disks with a gap of $G=20$ nm. The inset shows the same structure at a higher magnification. (Right) Non-interacting disks with $G = D + 40$ nm. The major symmetry axes of the patterned square lattices, [10] and [11] are also denoted.}
\label{fig1}
\end{center}
\end{figure}

\hspace{\parindent}
 Two sets of samples were prepared: one set for photo-emission electron microscopy (PEEM) studies employing x-ray magnetic circular dichroism (PEEM-XMCD) and the other set for magneto-optical Kerr effect (MOKE) measurements. The samples were prepared depositing elemental Fe (99.95 at\%) and Pd (99.95 at\%) in an ultra-high vacuum (base pressure below $\sim 2 \times 10^{-7}$ Pa) DC magnetron sputtering system, operating using high purity Argon gas (99.995 \%). The following sample structure was used: fused silica/Pd [40 nm]/Fe$_{13}$Pd$_{87}$ [10 nm]/Pd [2 nm]  \cite{ostman_hysteresis-free_2014}. After growth, electron beam lithography (EBL) was used to pattern the Fe$_{13}$Pd$_{87}$ layers into circular islands arranged on square lattices (see Fig.~\ref{fig1}). The Ar$^+$-ion milling process following development of the EBL resist and subsequent mask deposition, was stopped prior to penetration through the Pd seed layer, providing this way electrical continuity across the whole sample surface. The PEEM-XMCD samples with a disk diameter of $D = [75, 150, 350]$ nm were fabricated having two inter-disk distances; one with the nearest neighbour distance set to G = D + 40 nm, rendering the inter-disk coupling sufficiently weak to be ignored, and the other with an edge-to-edge gap of G = 20 nm to promote interactions between the magnetic elements \cite{guslienko_coupling_2001}. The MOKE samples were patterned into disks with a diameter of D~=~$[250, 350, 450]$ nm with a small edge-to-edge gap of G = 20 nm.

\subsection{PEEM-XMCD}
\hspace{\parindent}
To determine the magnetic state of the non-interacting and strongly coupled disks, photoemission electron microscopy, employing x-ray magnetic circular dichroism (PEEM-XMCD), was performed at the HERMES beamline at the SOLEIL synchrotron \cite{belkhou_hermes_2015}, and the 11.0.1 beamline at the Advanced Light Source \cite{doran_cryogenic_2012}. Prior to imaging, the samples were cooled from room temperature to roughly 100 K, in absence of any external magnetic field. The energy of the synchrotron beam was tuned to the Fe L$_3$-edge (708.4 eV), and magnetic contrast was obtained from the asymmetry ratio of intensities between right- and left-handed circularly polarised synchrotron radiation.

\subsection{Micromagnetic simulations and topological considerations}
\hspace{\parindent}
To understand the role of the interactions, we chose to explore the energies of the vortex and collinear state for different disk sizes, using the MuMax$^3$ micromagnetic simulation software \cite{vansteenkiste_design_2014}. The simulations, used to mimic the experimental conditions, were performed using one single disk, as well as square arrays of disks ($G = 20$ nm), with periodic boundary conditions. The saturation magnetisation, $M_{\text{sat}} = 3.5 \cdot 10^{5}$ A/M and exchange stiffness, $A_{\text{ex}} = 3.36 \cdot 10^{-12}$ J/m were chosen based on previous work by Östman et al. \cite{ostman_hysteresis-free_2014} and Ciuciulkaite et al. \cite{ciuciulkaite_collective_2019}. The in-plane cell size was set to 0.50(1)$l_{\text{ex}}$, where $l_{\text{ex}}$ is the exchange length as defined by $M_{\text{sat}}$ and $A_{\text{ex}}$ \cite{vansteenkiste_design_2014}.

\subsection{Hysteresis protocols}\label{hysprot}
\hspace{\parindent}
The magnetisation data, displayed in Fig.~\ref{Fig5}, was collected using a MOKE system in longitudinal configuration with the sample mounted on a cryostat, in a temperature range of 80 K $< T <$ 400 K. The $p$-polarised incident laser beam with $\lambda =$ 659 nm has a Gaussian profile with a spot diameter of roughly 2 mm. The number of islands contributing to the signal are therefore in the order of $10^7$. The reflected laser beam was passed through an analyser (extinction ratio of $10^5:1$) and then captured using a Si biased detector, connected to a pre-amplifier and lock-in amplifier. In addition, the lock-in amplifier was used to modulate the incident laser beam using a Faraday cell. The sinusoidal external magnetic field, was applied along [10] of the square lattice (see Fig.~\ref{fig1}), had an amplitude of 40 mT and a frequency of 0.11 Hz unless otherwise stated. The magnetisation data in all figures have been binned (3:1) for aesthetic purposes.

\section{RESULTS AND DISCUSSION}\label{res}
\begin{figure}[t!]
\begin{center}
\includegraphics[width=1\linewidth]{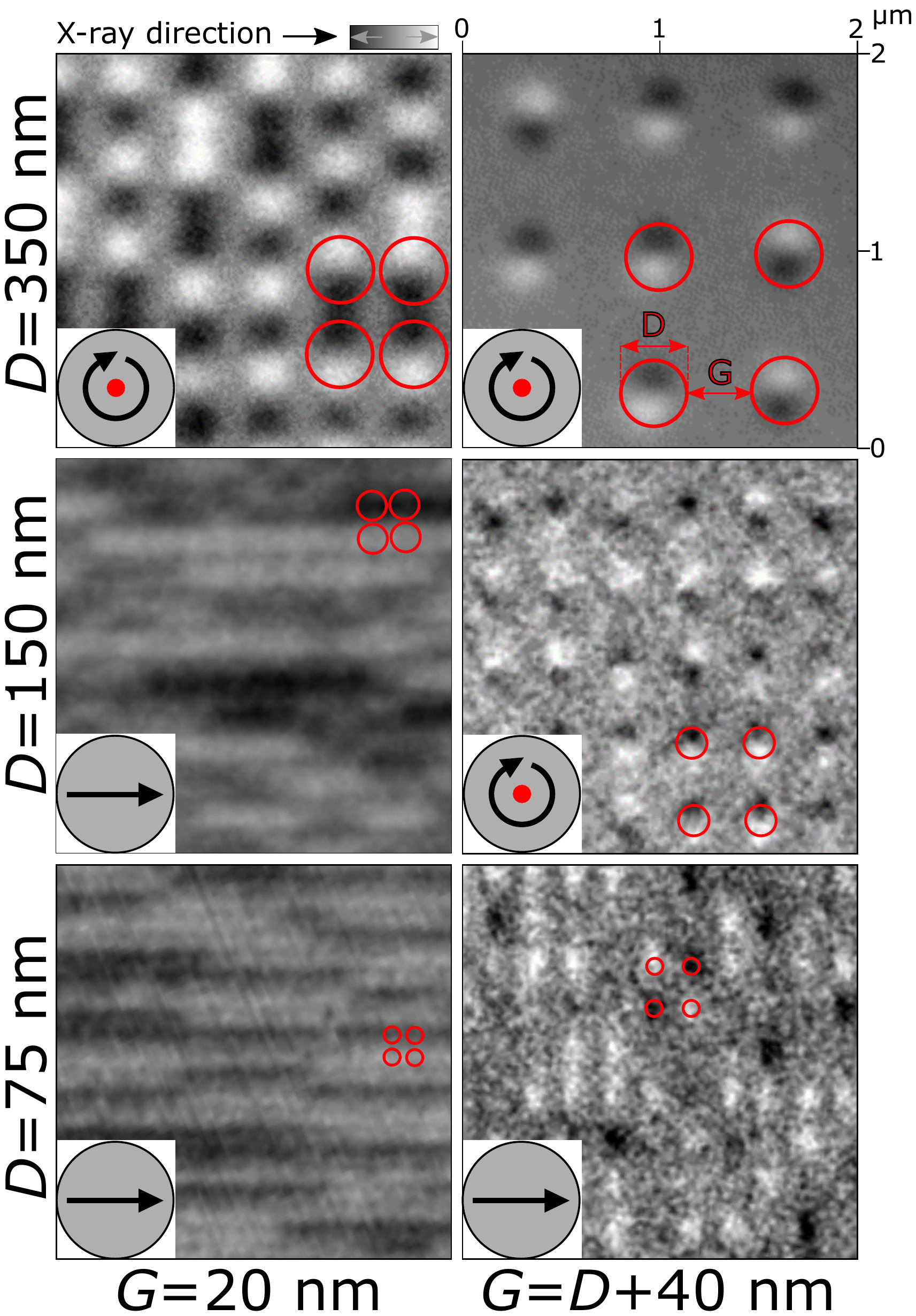}
\caption{Diameter and interaction dependence of the disk magnetic texture. PEEM-XMCD images of interacting islands (left column), and non-interacting islands (right column), recorded at approximately 100 K after cooling from room temperature in absence of external fields. Disks with a diameter of 350 nm have a preferred vortex texture for both interacting to non-interacting disks, while disks with 75 nm diameter end up in the collinear state in both cases. Disks with a diameter of 150 nm, display a stabilisation of the collinear state when the distance between the islands is small, while otherwise exhibiting vortex textures. The red circles indicate the disk sizes and positions, obtained by overlapping the PEEM-X-ray absorption spectroscopy images.}
\label{fig2}
\end{center}
\end{figure}

\hspace{\parindent}
We begin by discussing the size dependence of internal spin textures and spontaneous magnetic order in circular islands. To this end, we fabricated samples with nominal disk diameters $D = [75, 150, 350]$ nm, having edge-to-edge gaps $G=[20, D+40]$ nm, on one and the same wafer. This approach allows us to disregard any uncertainty related to $e.g.$ composition and thickness of the islands. Representative parts of the samples with 150 nm islands are illustrated in Fig. \ref{fig1}. The results of photo-emission electron microscopy on these samples, with magnetic contrast obtained by using x-ray magnetic circular dichroism (PEEM-XMCD), are displayed in Fig. \ref{fig2}. As inferred by the black and white contrast, the disks with a diameter of 75 nm are all found to exhibit a collinear component for both small and large distances between the islands. Similarly, the 350 nm islands are found to be in a vortex state for both the short and long distance between the islands, with no preferred sense of rotation. When the disk diameter is 150 nm, a vortex state is obtained when the distance between the islands is large, while a substantial amount of mesospins have a collinear component, when the distance between the islands is 20 nm. Thus, for this particular diameter of islands, the interaction appears to be strong enough to influence the inner magnetic states. 
To understand the role of the interactions and the interplay between the involved length-scales, we need to have a look at the energy landscape of the magnetic textures.

\begin{figure}[t!]
\begin{center}
\includegraphics[width=1\linewidth]{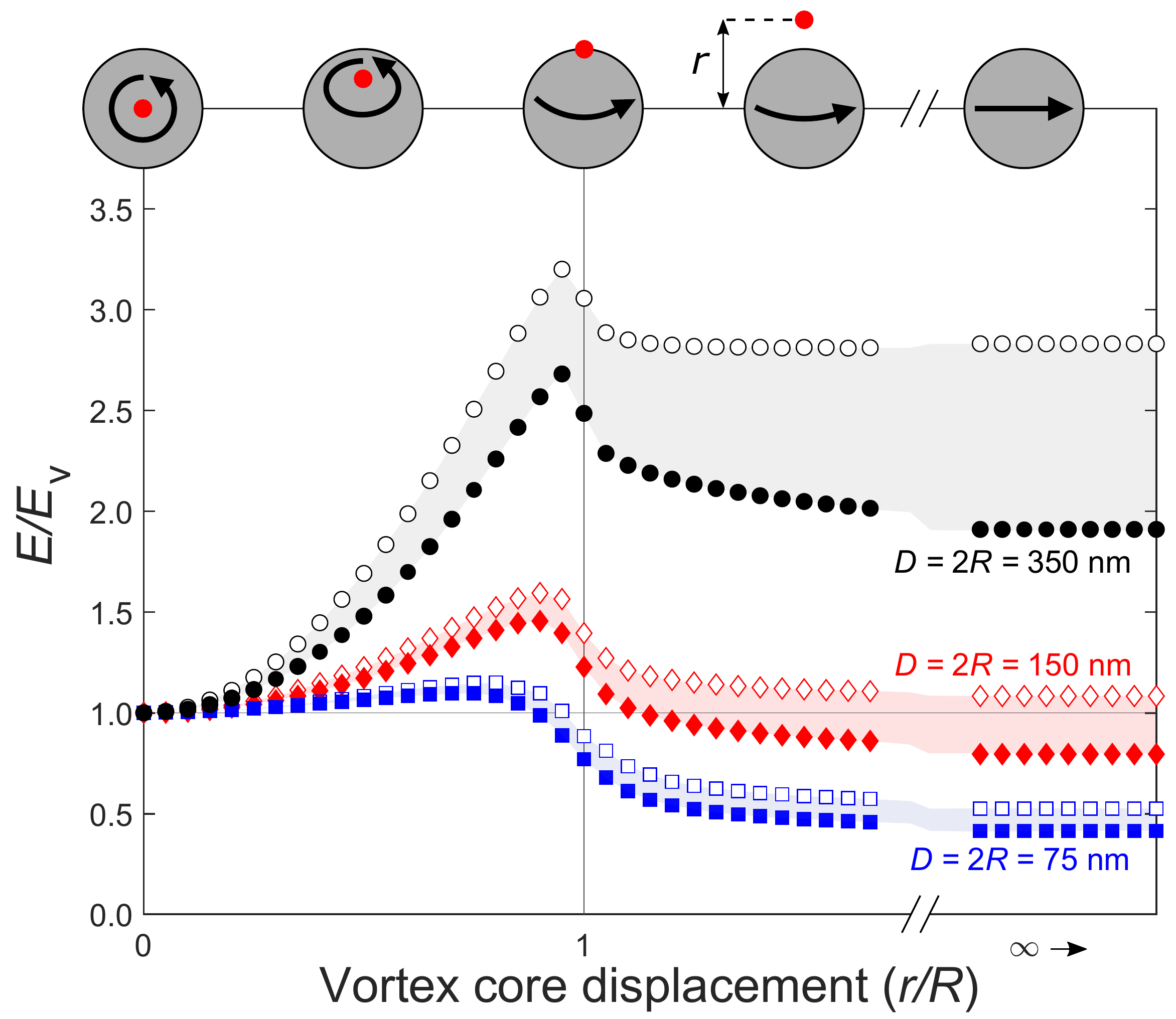}
\caption{The energy landscape of the transition between the vortex to the collinear state. The landscape was obtained numerically by gradually moving the vortex core outwards from the centre of the disks. Filled symbols represent the energy values of interacting disks ($G$ = 20 nm), while empty symbols for non-interacting disks. The shaded areas represent the magnetostatic coupling $E_{\text{j}}$. The energies plotted are normalised to the total energy of the vortex state $E_{\text{v}}$. The energy barrier between the two states is tunable by choice of disk diameter and inter-island interaction strength. The top part shows a schematic of the states for different values of $r$, where the red dot indicates the position of the vortex core, when $r > R$, a $C$-state is obtained.}
\label{fig3}
\end{center}
\end{figure}

The magnetic energy of the metamaterial can be expressed as $E = E_{\text{t}}+E_{\text{s}}+E_{\text{j}}$, where $E_{\text{t}}$ is the energy cost of the magnetic texture arising from exchange interactions within the islands, $E_{\text{s}}$ is the magnetostatic energy and $E_{\text{j}}$ is the energy associated with the magnetostatic coupling between the islands. By calculating the energy as a function of the vortex core displacement $r$ for a disk with radius $R$, a single path -- out of many -- in the energy landscape separating the collinear and the vortex state can be obtained. Ding et al. used the same reasoning when calculating the energy barrier separating the two states in a single Co dot analytically \cite{ding_magnetic_2005}. Here, we chose to do it numerically as the calculations can be generalised to include interactions of elements in an arbitrary array. The results obtained from calculations of interacting ($G = 20$ nm) and non-interacting ($G = D + 40$ nm) disks with $D = [75, 150, 350]$ nm are shown in Fig. \ref{fig3}.
When a vortex core is at the centre of an island, its displacement is defined to be zero and the energy is $E \approx E_{\text{t}}$ because $E_{\text{s}} \approx E_{\text{j}} \approx 0$. Shifting the core from the centre gives rise to a collinear component, and consequently a stray field with a corresponding magnetostatic energy. For the purpose of the calculation, the (virtual) vortex core can even be moved outside the disk ($r>R$), corresponding to a $C$-state (see Fig. \ref{fig3}), with varying degree of gradients in the magnetic texture. 
The energy maxima obtained at $r \approx 0.9 R$ can be viewed as activation barriers separating the vortex and the collinear states.
The energy difference between the interacting and non-interacting islands 
increases with increasing $r$, as illustrated in the figure. 

In these simulations we use the energies obtained by MuMax$^3$'s built in functions for the vortex and the collinear state, without allowing for relaxation of the magnetisation with respect to the total energy. Using this approach, the intermediate states can be calculated without the systems collapsing into either the vortex or the collinear state, enabling us to estimate the height of the activation barrier. Qualitatively, the same results can be obtained in relaxed systems by moving the core by means of an applied magnetic field. This approach, however, does not provide any information on the potential in the vicinity of the maximum of the activation barrier. In the simulations with interacting islands, all vortex cores were displaced such that the collinear component of each island was along [10], mimicking the influence of an applied field along [10] as in the MOKE experiments. The choice of either uniform or alternating vortex chirality had a negligible impact on the outcome of the simulations.

\begin{figure}[t!]
\begin{center}
\includegraphics[width=1\linewidth]{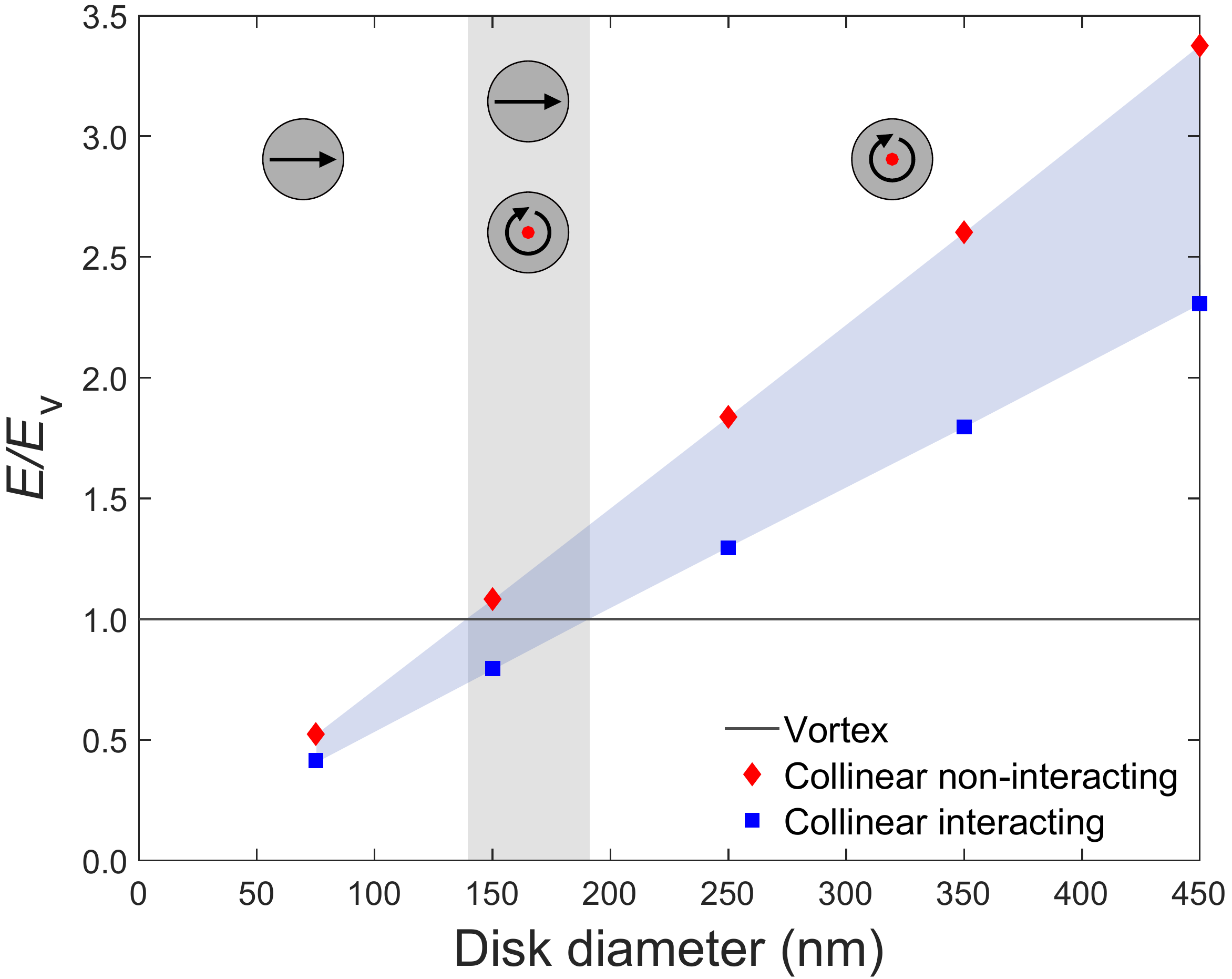}
\caption{The boundary between vortex and collinear states. The energy of the collinear state is lowered (blue squares) by inter-island interactions. As a result, it is possible to stabilise collinear states for a range of diameters which otherwise favour formation of vortices (gray shaded area). The energies plotted are normalised to the total energy of the vortex state $E_{\text{v}}$.}
\label{fig4}
\end{center}
\end{figure}

The size dependence in the energy of interacting and non interacting mesospins is summarised in Fig. \ref{fig4}. For non-interacting islands the collinear state has the lowest energy when $D \lesssim 140$ nm, while a vortex state is favoured when the diameter is larger. With an inter-island distance of 20 nm, a collinear state is favoured up to about $D\approx 190$ nm. Consequently the critical size at which the vortex state is favoured is shifted to larger diameters when interactions become prominent. Hence, a region of bistability (marked by a grey shading in the figure) is obtained. We have thereby rationalised the results displayed in Fig. \ref{fig2}, ${i.e.}$ why the 150 nm islands form vortices in absence of interactions, while a significant amount of the mesospins show a collinear component when they interact.
When collinear, the mesospins can be viewed as being two-dimensional \cite{Arnalds_XY} (XY-rotor) and zero-dimensional \cite{ostman_hysteresis-free_2014} when in a vortex state. At the same time, the magnetic texture can be defined in terms of topological quasiparticles \cite{Zhang:2015de, Donnelly_2020fh}, as $e.g.$ illustrated in Fig. \ref{Fig6} and listed in Table~1 (Appendix~\ref{mesoclass}). The states displayed in Fig. \ref{Fig6} can thereby be defined as having topological charge of +1, determined by $1-g$, with $g$ being the genus of the structure \cite{Tchernyshyov:2005gs}. The boundary of the islands can host magnetic charges with fractional winding numbers, $w$, which together with the winding number of the bulk charge, $q$, must add up to 1 (see Fig. \ref{Fig6}) \cite{Tchernyshyov:2005gs, Sloetjes:2020iw}. Consequently, the winding of the vortex can been seen as being transferred to the edge of the island when the magnetisation changes from a vortex to a collinear state ( $r > R$).  
The transition from a vortex to a collinear state therefore involves no change in the total winding number, while the magnetic cores and their polarity annihilate at the edges of the islands.

\begin{figure}[t!]
\begin{center}
\includegraphics[width=1\linewidth]{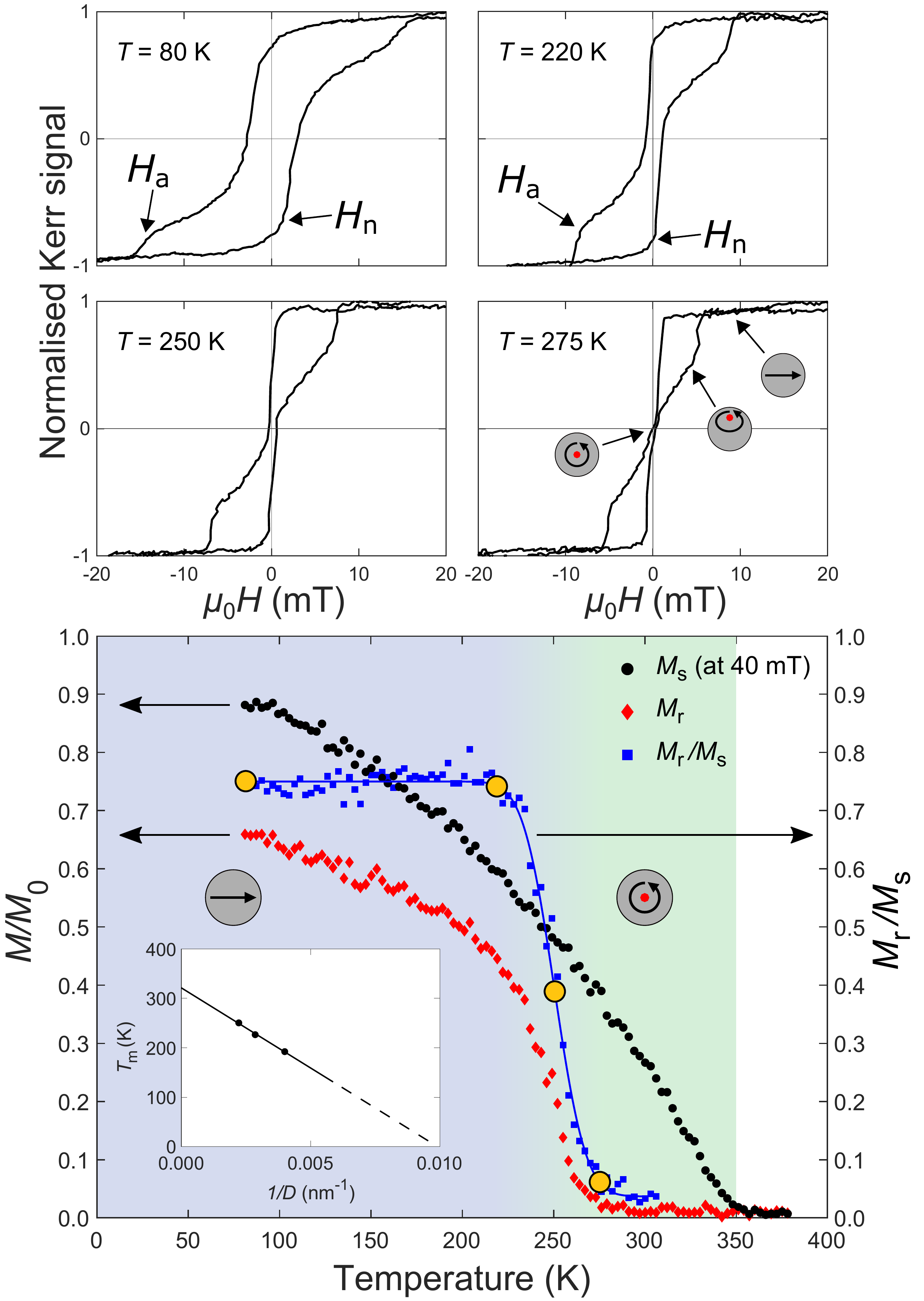}
\caption{Temperature dependence of the dimensionality transition in a magnetic metamaterial. Magnetisation measurements with the applied field along [10] (see Fig. \ref{fig1}) for the 450 nm islands having a gap of $G$~=~20~nm. (Top) Representative hysteresis curves at 80, 220, 250 and 275 K. The curves have been normalised for each individual temperature. The nucleation field crosses $\mu_0 H = 0$ at approximately 220 K, indicating a stability of the collinear state below this temperature, during the timescale of the measurement. (Bottom) The magnetisation as a function of temperature at 0 and 40 mT as well as the ratio $M_{\text{m}}$=$M_{\text{r}}$/$M_{\text{s}}$. The orange filled circles mark the temperatures at which the hysteresis curves, shown above, were taken. The size dependence of the transition temperature, $T_{\text{m}}$, for island sizes $D = [250, 350, 450]$ nm is shown in the inset. The error bars of the data in the inset are smaller than the data point symbols.}
\label{Fig5}
\end{center}
\end{figure}

While it is trivial to annihilate vortices by applying an external field, the opposite is certainly not true. For this reason we chose to focus on the field and temperature dependence of islands with a diameter of 250 nm and larger ($D = [250, 350, 450]$ nm), with gaps $G$ = 20 nm. Disks in this size range spontaneously form vortices when cooled, while the application of an external magnetic field results in a collinear state, allowing us to control the magnetic texture of the mesospins. The stability of the $dressed$ collinear state can thus be investigated by removing the external field while monitoring the magnetisation of the samples. Representative magnetisation loops for the 450 nm islands, recorded at four different temperatures, are provided in the top half of Fig. \ref{Fig5}. The bottom part illustrates the temperature dependence of the magnetisation.

 At $T$ = 275 K the hysteresis loops display a typical vortex nucleation ($H_{\text{n}}$) and annihilation ($H_{\text{a}}$) signature with zero remanence \cite{cowburn_single-domain_1999}, in line with both the PEEM-XMCD results as well as the simulations. At temperatures below $\approx$ 250 K, the sample exhibits clear remanence, consistent with the presence of a ferromagnetic state. Ferromagnetic response requires alignment of mesospins with a net moment, in stark contrast to the zero field PEEM-XMCD results. The reduction of the remanent magnetisation ($M_{\text{r}}$) with temperature contains both the decrease of the moment of the material (intrinsic material properties) as well as changes in the texture and orientation of the mesospins. To disentangle these contributions we need to identify their signatures. Changes in the material-related magnetisation can be described by a power law up to the ordering temperature of the materials ($T_{\mathrm{C}}$), while the temperature dependence of the texture and orientation of the mesospins is unknown. However, separation of the two contributions can be obtained by identifying the difference in their field dependence. For instance, it is sufficient to apply a relatively weak field to remove most of the magnetic texture within a disk, while weak fields only marginally affect the thermally induced excitations of the magnetisation in the material. This is seen in the field dependence displayed in Fig.~\ref{Fig5}: at $T$ = 275 K, a field of approximately 6 mT is sufficient for obtaining a transition from a vortex to a collinear state. Nevertheless, this field does little to alter the thermally induced reduction of the magnetisation. We can therefore consider the magnetisation at a field of 40 mT as predominantly representative for the material, $i.e.$ we define this magnetisation as the saturation of the mesoscopic texture ($M_{\text{s}}$). Consequently, the temperature dependence of $M_{\text{m}}$=$M_{\text{r}}$/$M_{\text{s}}$ can be thought of as corresponding to the temperature dependence of the remanence of the mesospin texture. 
 
The results displayed in the bottom half of Fig.~\ref{Fig5} (450 nm islands), illustrate $M_{\text{r}}$, $M_{\text{s}}$ and $M_{\text{m}}$=$M_{\text{r}}$/$M_{\text{s}}$, the inferred remanence of the mesospins. A plateau with $M_{\text{m}} = 3/4$ is observed, consistent with constant magnetic texture and thereby a fixed topology of the mesospins, below 220K. At the same time, a substantial fraction of the moment (1/4) is perpendicular and/or antiparallel with respect to the direction of the net magnetisation. At 220 K there is an abrupt change in $M_{\text{m}}$, which vanishes at 270 K. From 270 K and up to the intrinsic Curie temperature of the material, the obtained hysteresis loops are consistent with the presence of vortex states. Thus, we observe a sharp transition from a collective state of interacting mesospins to non-interacting vortex states of the elements with zero net magnetisation. To this observation we assign a collapse from a state with a spin dimensionality of 2, to a state with zero spin dimensionality. 
The temperature dependence of this transition can be determined by fitting an error function to the data, from which the temperature $T_{\text{m}}$ where $M_{\text{m}}(T_{\text{m}})=\frac{1}{2}M_{\text{m}}(T=0~\mathrm{K})$ (inset of Fig. \ref{Fig5}, bottom panel) is determined. An error function was chosen without assigning any physical interpretation to it. A clear size dependence ($T_{\text{m}} \propto 1-1/D$) is observed. The results imply an increase in stability of the collinear state with increasing disk size. Extrapolation to infinite size yields a transition temperature approaching $T_{\text{C}}$ of the continuous film, as expected. Extrapolation to zero in $T_{\text{m}}$ yields $\approx$ 100 nm island size, which corresponds to the lower limit for the possibility to observe the transition.
However, as illustrated in Fig. \ref{fig4}, the collinear state is already 
favoured when $D \lesssim 190$ nm, rendering this region unattainable. 
As the vortex state is energetically favoured above this size, the observed transitions in 250, 350 and 450 nm elements are kinetically limited and do not correspond to thermodynamic phase transitions in a strict sense. This is further emphasised by the observed frequency dependence of the transition temperature, $T_{\text{m}}$ (see Appendix~\ref{kinetics}). The dressed mesospins (collinear state) in this size range, are therefore in an arrested metastable state with the boundaries of the mesospins being vital for the creation and annihilation process of the vortex cores.


In conclusion, the temperature dependence of the topologically homeomorphic transition is not only defined by the material properties and size of the mesospins but also by their mutual interactions and internal degrees of freedom. In particular, the distance between the elements alters the inter-island interactions and thereby the internal magnetic states, while at the same time, the internal magnetic states affect the interactions and thereby the magnetic order. The interplay between these two length-scales leads to an exotic transition, that is to say that it is not based on thermally induced randomisation. Rather, the transition pertains to a thermally activated change within the elements, resulting in a collapse of the effective interaction strength and the mesospin dimensionality. The transition discussed here does not have any trivial classical counterpart, calling for a new conceptual framework involving mutual dependence of energy and length-scales \cite{Wilson:1979wn}. The results represent stumbling steps towards understanding emergent properties and complexity that may extend beyond the immediate field of physics \cite{castellano2009statistical,michaud2018social}. 

\section*{Data availability}

The data that support the ﬁndings of this study are available from the authors upon reasonable request.

\section*{Acknowledgments}

The authors would like to acknowledge the excellent user support provided to them at the Advanced Light Source (Dr. Andreas Scholl, Dr. Rajesh Chopdekar) and SOLEIL (Dr. Rahid Belkhou) synchrotrons, as well as Dr. Erik Östman for helping out with data collection. This research used resources of the Advanced Light Source, a U.S. DOE Office of Science User Facility under contract no. DE-AC02-05CH11231. The excellent support and infrastructure of the MyFab facility at the \AA ngstr\"om Laboratory of Uppsala University is also highly appreciated. B.H. acknowledges financial support from the Swedish Research Council and the Swedish Foundation for Strategic Research. V.K. acknowledges support from the Knut and Alice Wallenberg Foundation project ``{\it Harnessing light and spins through plasmons at the nanoscale}'' (2015.0060) and the Swedish Research Council (Project No. 2019-03581).



\appendix

\section{Mesospin classification}\label{mesoclass}

\begin{figure}[h!]
    \centering
    \includegraphics[width=1\linewidth]{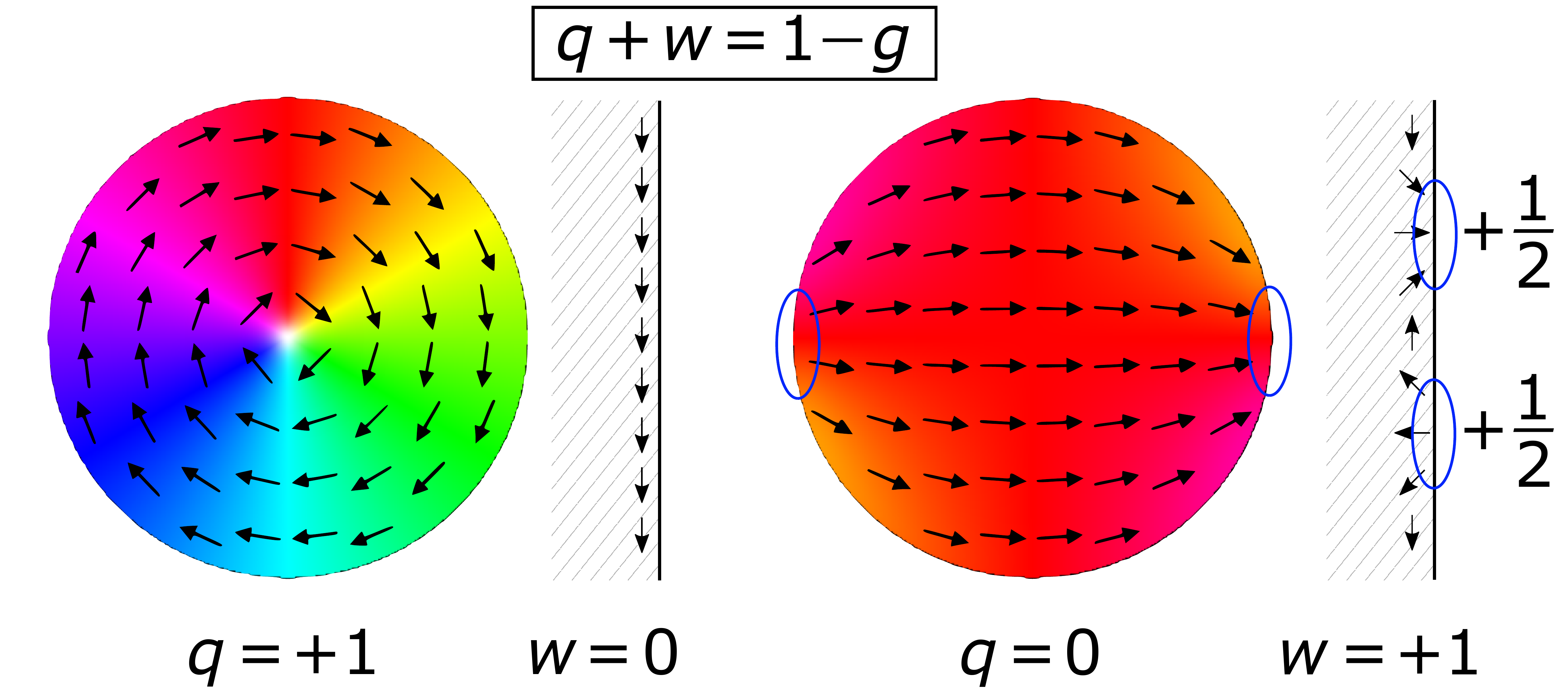}
    \caption{Magnetic texture and topology of a vortex (left) and a collinear state (right). There is substantial texture in the magnetic state referred to as collinear, however, the net transverse moment is zero. The interior topological charge $q = +1$ (vortex) can be transferred to two topological charges, with $w = +\frac{1}{2}$, at the edges (collinear)  \cite{Tchernyshyov:2005gs}. The schematics next to the textures depict the changes in the direction of the magnetisation with respect to the edge and the resulting winding number.} 
    \label{Fig6}
\end{figure}

Fig. \ref{Fig6} illustrates the vortex and the collinear states, as well as their corresponding topological charges residing in the bulk and on the edges respectively. A transition between the two states is topologically allowed for elements of finite size, and can be driven by applying external fields or thermally induced fluctuations \cite{ostman_hysteresis-free_2014}. The amount of topological charge that can be residing in an element is determined by the genus $g$ (number of holes) of the structure, providing an additional knob for the design of texture topology in magnetic metamaterials. This transition is followed by a change in the effective spin dimensionality of the mesospins, which is 0 for the vortex and 2 for the collinear state (see also Table~1). Note that in the interior of the mesospins, non-integer winding numbers cannot exist \cite{Tchernyshyov:2005gs, Zhang:2015de}. This is not the case though for the edges where half-integer winding numbers are possible and thus topological charges, $w$, which further correlate to stray field emanating from or into the mesospin \cite{Tchernyshyov:2005gs}. These appear in pairs, resulting in integer numbers for the total charge on the edge.

The spin dimensionality $d$ relates to the behaviour of the effective mesospin moment, which for planar mesospins can be 0-, 1- or 2-dimensional \cite{shinjo_magnetic_2000,ostman_ising-like_2018,Arnalds_XY}, as illustrated in Table~\ref{Tab1}. The sense of rotation $c$ describes the circulation (clock- and anticlock-wise) of the magnetic texture and acquires non-zero values in the vortex case (0-dimensional mesospin) \cite{Jenkins_vortices_2014}. The polarity $p$ relates to the direction of the out-of-plane component (up or down) of the magnetisation for the core in a vortex. Finally, when considering the planar topology, the bulk topological charge $q$, defines the magnetic texture of the mesospins, being +1 for vortex states and 0 for Ising or XY-rotor states. The blue coloured cells refer to mesospin properties and their associated transitions, particularly studied in this work. \\

\begin{table}[t!]
    \centering
    \caption{Classification scheme for the planar mesospins.}
    \includegraphics[width=1\linewidth]{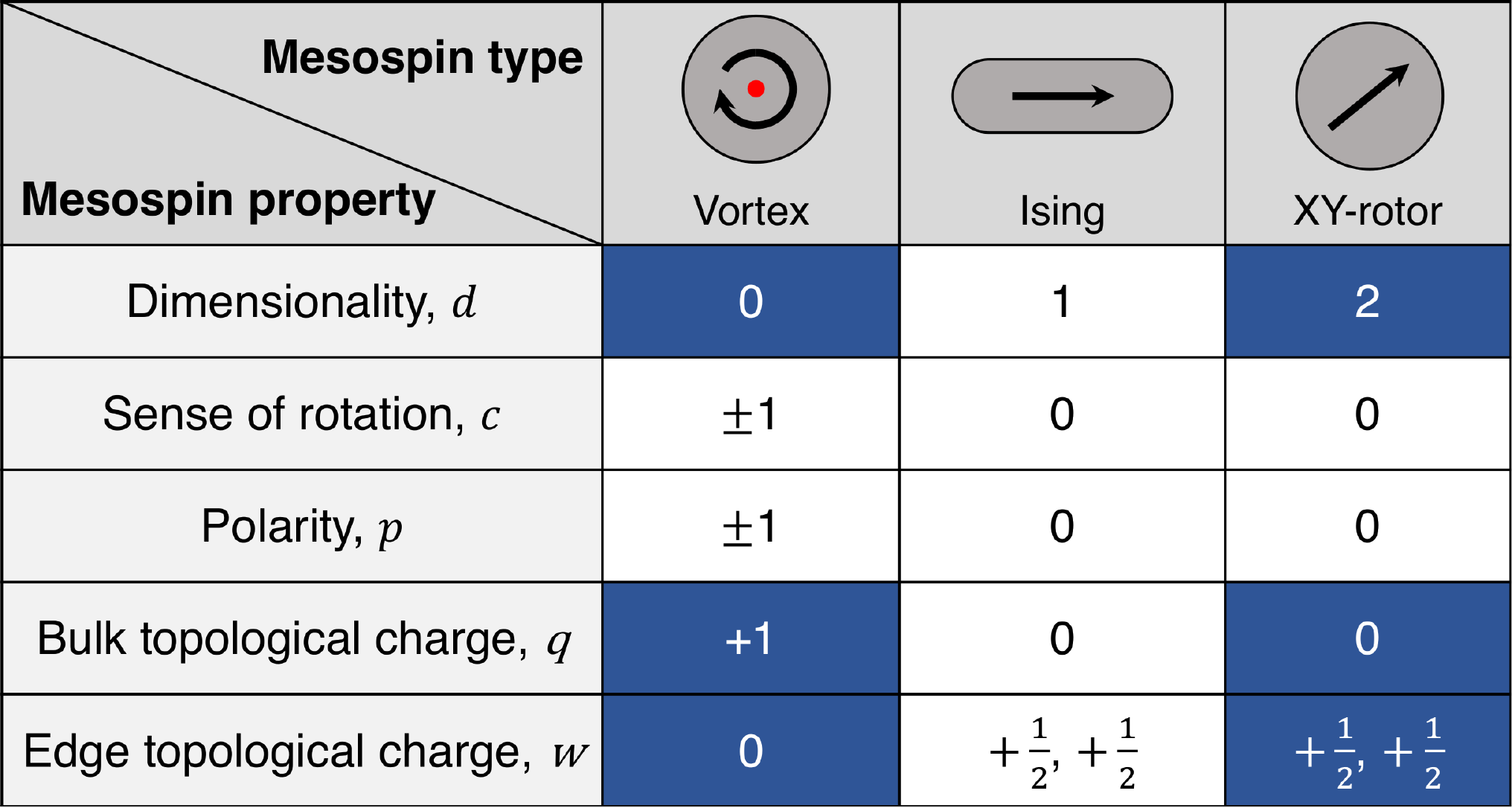}
    \label{Tab1}
\end{table}

\section{Kinetics}\label{kinetics}

\begin{figure}[h!]
\begin{center}
\includegraphics[width=0.8\linewidth]{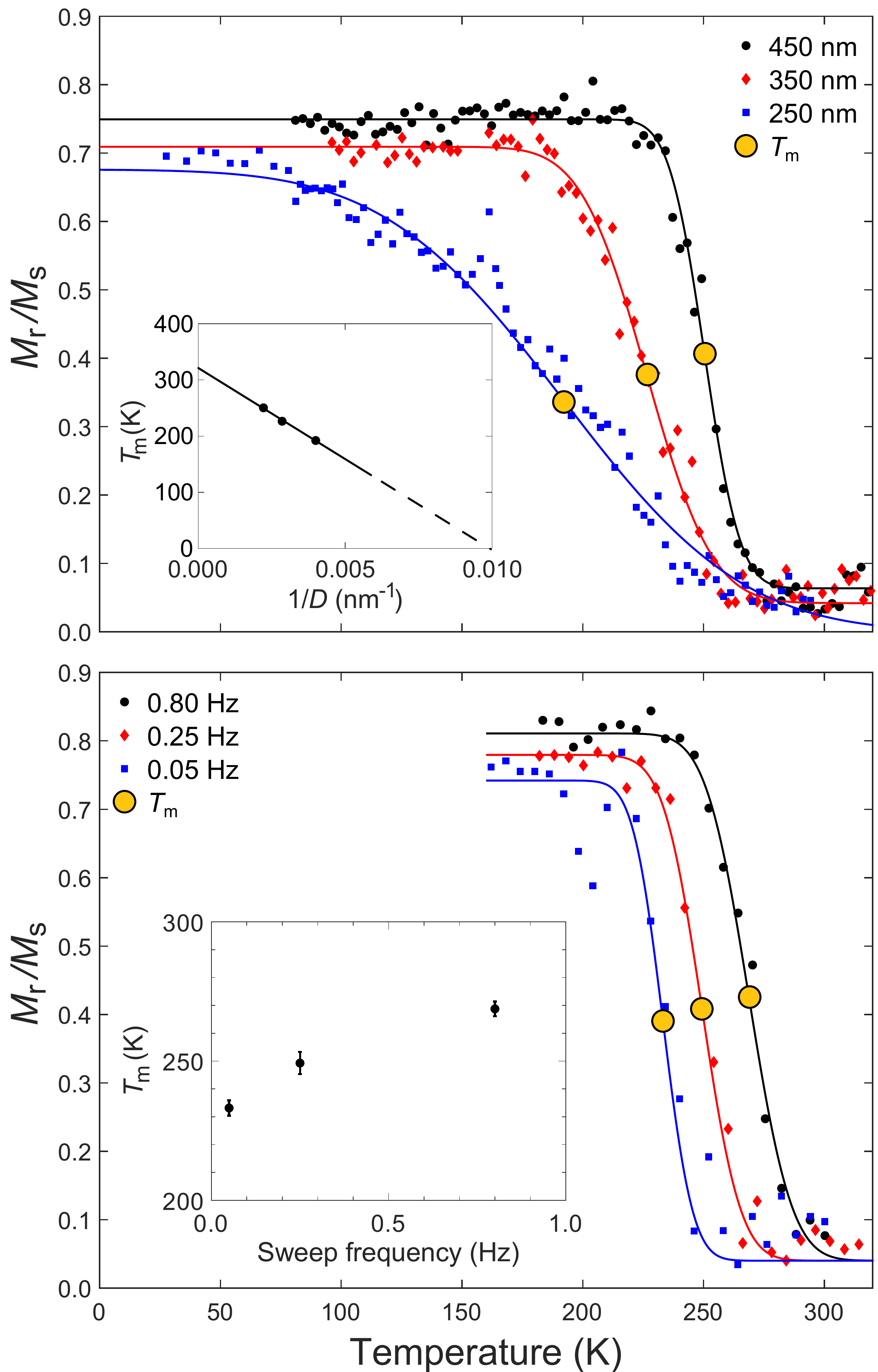}
\caption{Size and frequency dependence of the dimensionality transition. (Top) Temperature dependence of the remanent versus the in-field (40 mT) magnetization ratio, for all available disk diameters and for a gap $G$ = 20 nm. A linear scaling of the transition temperature $T_{\text{m}}$, determined from the inflection points of the fitted curves versus the reciprocal of the disk diameter $D$, can been seen in the inset. (Bottom) The frequency dependence of the transition for a disk diameter $D$ = 450 nm and gap $G$ = 20 nm. A monotonic shift to higher temperature with increasing frequency is observed, compatible with thermally activated relaxation processes of magnetic metamaterials
. The orange filled circles in each panel, mark the extracted temperatures presented in the respective insets.}
\label{fig:Mrs_size}
\end{center}
\end{figure}

To obtain a better understanding on the interplay between kinetics and thermodynamics, we explored the influence of the frequency of the applied field on the obtained remanent magnetisation. The dwell time at zero field provides a time window for a relaxation, which is proportional to the inverse scanning frequency. The findings are illustrated in the lower panel of Fig. \ref{fig:Mrs_size}, depicting a clear frequency dependence of the response from the 450 nm disks. Increasing the scanning frequency, results in an increase of the transition temperatures (inset in Fig. \ref{fig:Mrs_size}, lower panel), illustrating the importance of the kinetic limitations on the transition \cite{Pohlit_PRB_2020,Sloetjes_arXiv_2020}. Consequently, we conclude that the ferromagnetic component observed at low temperatures, corresponds to a metastable arrested state. The frequency dependence can not be captured by a single exponential, highlighting the non-Arrhenius behaviour of the homeomorphic transition.

\begin{thebibliography}{35}%
\makeatletter
\providecommand \@ifxundefined [1]{%
 \@ifx{#1\undefined}
}%
\providecommand \@ifnum [1]{%
 \ifnum #1\expandafter \@firstoftwo
 \else \expandafter \@secondoftwo
 \fi
}%
\providecommand \@ifx [1]{%
 \ifx #1\expandafter \@firstoftwo
 \else \expandafter \@secondoftwo
 \fi
}%
\providecommand \natexlab [1]{#1}%
\providecommand \enquote  [1]{``#1''}%
\providecommand \bibnamefont  [1]{#1}%
\providecommand \bibfnamefont [1]{#1}%
\providecommand \citenamefont [1]{#1}%
\providecommand \href@noop [0]{\@secondoftwo}%
\providecommand \href [0]{\begingroup \@sanitize@url \@href}%
\providecommand \@href[1]{\@@startlink{#1}\@@href}%
\providecommand \@@href[1]{\endgroup#1\@@endlink}%
\providecommand \@sanitize@url [0]{\catcode `\\12\catcode `\$12\catcode
  `\&12\catcode `\#12\catcode `\^12\catcode `\_12\catcode `\%12\relax}%
\providecommand \@@startlink[1]{}%
\providecommand \@@endlink[0]{}%
\providecommand \url  [0]{\begingroup\@sanitize@url \@url }%
\providecommand \@url [1]{\endgroup\@href {#1}{\urlprefix }}%
\providecommand \urlprefix  [0]{URL }%
\providecommand \Eprint [0]{\href }%
\providecommand \doibase [0]{http://dx.doi.org/}%
\providecommand \selectlanguage [0]{\@gobble}%
\providecommand \bibinfo  [0]{\@secondoftwo}%
\providecommand \bibfield  [0]{\@secondoftwo}%
\providecommand \translation [1]{[#1]}%
\providecommand \BibitemOpen [0]{}%
\providecommand \bibitemStop [0]{}%
\providecommand \bibitemNoStop [0]{.\EOS\space}%
\providecommand \EOS [0]{\spacefactor3000\relax}%
\providecommand \BibitemShut  [1]{\csname bibitem#1\endcsname}%
\let\auto@bib@innerbib\@empty
\bibitem [{\citenamefont {Heyderman}\ and\ \citenamefont
  {Stamps}(2013)}]{Heyderman_Review_2013}%
  \BibitemOpen
  \bibfield  {author} {\bibinfo {author} {\bibfnamefont {L.~J.}\ \bibnamefont
  {Heyderman}}\ and\ \bibinfo {author} {\bibfnamefont {R.~L.}\ \bibnamefont
  {Stamps}},\ }\href {\doibase 10.1088/0953-8984/25/36/363201} {\bibfield
  {journal} {\bibinfo  {journal} {Journal of Physics: Condensed Matter}\
  }\textbf {\bibinfo {volume} {25}},\ \bibinfo {pages} {363201} (\bibinfo
  {year} {2013})}\BibitemShut {NoStop}%
\bibitem [{\citenamefont {{\"O}stman}\ \emph
  {et~al.}(2018{\natexlab{a}})\citenamefont {{\"O}stman}, \citenamefont
  {Arnalds}, \citenamefont {Kapaklis}, \citenamefont {Taroni},\ and\
  \citenamefont {Hj{\"o}rvarsson}}]{ostman_ising-like_2018}%
  \BibitemOpen
  \bibfield  {author} {\bibinfo {author} {\bibfnamefont {E.}~\bibnamefont
  {{\"O}stman}}, \bibinfo {author} {\bibfnamefont {U.~B.}\ \bibnamefont
  {Arnalds}}, \bibinfo {author} {\bibfnamefont {V.}~\bibnamefont {Kapaklis}},
  \bibinfo {author} {\bibfnamefont {A.}~\bibnamefont {Taroni}}, \ and\ \bibinfo
  {author} {\bibfnamefont {B.}~\bibnamefont {Hj{\"o}rvarsson}},\ }\href
  {\doibase 10.1088/1361-648X/aad0c1} {\bibfield  {journal} {\bibinfo
  {journal} {Journal of Physics: Condensed Matter}\ }\textbf {\bibinfo {volume}
  {30}},\ \bibinfo {pages} {365301} (\bibinfo {year}
  {2018}{\natexlab{a}})}\BibitemShut {NoStop}%
\bibitem [{\citenamefont {Nisoli}\ \emph {et~al.}(2013)\citenamefont {Nisoli},
  \citenamefont {Moessner},\ and\ \citenamefont {Schiffer}}]{Nisoli_2013}%
  \BibitemOpen
  \bibfield  {author} {\bibinfo {author} {\bibfnamefont {C.}~\bibnamefont
  {Nisoli}}, \bibinfo {author} {\bibfnamefont {R.}~\bibnamefont {Moessner}}, \
  and\ \bibinfo {author} {\bibfnamefont {P.}~\bibnamefont {Schiffer}},\ }\href
  {\doibase 10.1103/RevModPhys.85.1473} {\bibfield  {journal} {\bibinfo
  {journal} {Rev. Mod. Phys.}\ }\textbf {\bibinfo {volume} {85}},\ \bibinfo
  {pages} {1473} (\bibinfo {year} {2013})}\BibitemShut {NoStop}%
\bibitem [{\citenamefont {Nisoli}\ \emph {et~al.}(2017)\citenamefont {Nisoli},
  \citenamefont {Kapaklis},\ and\ \citenamefont {Schiffer}}]{Nisoli:2017hg}%
  \BibitemOpen
  \bibfield  {author} {\bibinfo {author} {\bibfnamefont {C.}~\bibnamefont
  {Nisoli}}, \bibinfo {author} {\bibfnamefont {V.}~\bibnamefont {Kapaklis}}, \
  and\ \bibinfo {author} {\bibfnamefont {P.}~\bibnamefont {Schiffer}},\ }\href
  {\doibase 10.1038/nphys4059} {\bibfield  {journal} {\bibinfo  {journal}
  {Nature Physics}\ }\textbf {\bibinfo {volume} {13}},\ \bibinfo {pages} {200}
  (\bibinfo {year} {2017})}\BibitemShut {NoStop}%
\bibitem [{\citenamefont {Rougemaille}\ and\ \citenamefont
  {Canals}(2019)}]{Rougemaille_2019}%
  \BibitemOpen
  \bibfield  {author} {\bibinfo {author} {\bibfnamefont {N.}~\bibnamefont
  {Rougemaille}}\ and\ \bibinfo {author} {\bibfnamefont {B.}~\bibnamefont
  {Canals}},\ }\href {\doibase 10.1140/ep jb/e2018-90346-7} {\bibfield
  {journal} {\bibinfo  {journal} {Eur. Phys. J. B}\ }\textbf {\bibinfo {volume}
  {92}},\ \bibinfo {pages} {62} (\bibinfo {year} {2019})}\BibitemShut {NoStop}%
\bibitem [{\citenamefont {Gliga}\ \emph {et~al.}(2013)\citenamefont {Gliga},
  \citenamefont {K{\'a}kay}, \citenamefont {Hertel},\ and\ \citenamefont
  {Heinonen}}]{Gliga_PRL_2013}%
  \BibitemOpen
  \bibfield  {author} {\bibinfo {author} {\bibfnamefont {S.}~\bibnamefont
  {Gliga}}, \bibinfo {author} {\bibfnamefont {A.}~\bibnamefont {K{\'a}kay}},
  \bibinfo {author} {\bibfnamefont {R.}~\bibnamefont {Hertel}}, \ and\ \bibinfo
  {author} {\bibfnamefont {O.~G.}\ \bibnamefont {Heinonen}},\ }\href {\doibase
  10.1103/PhysRevLett.110.117205} {\bibfield  {journal} {\bibinfo  {journal}
  {Physical Review Letters}\ }\textbf {\bibinfo {volume} {110}},\ \bibinfo
  {pages} {117205} (\bibinfo {year} {2013})}\BibitemShut {NoStop}%
\bibitem [{\citenamefont {Gliga}\ \emph {et~al.}(2015)\citenamefont {Gliga},
  \citenamefont {K{\'a}kay}, \citenamefont {Heyderman}, \citenamefont
  {Hertel},\ and\ \citenamefont {Heinonen}}]{Gliga_PRB_2015}%
  \BibitemOpen
  \bibfield  {author} {\bibinfo {author} {\bibfnamefont {S.}~\bibnamefont
  {Gliga}}, \bibinfo {author} {\bibfnamefont {A.}~\bibnamefont {K{\'a}kay}},
  \bibinfo {author} {\bibfnamefont {L.~J.}\ \bibnamefont {Heyderman}}, \bibinfo
  {author} {\bibfnamefont {R.}~\bibnamefont {Hertel}}, \ and\ \bibinfo {author}
  {\bibfnamefont {O.~G.}\ \bibnamefont {Heinonen}},\ }\href {\doibase
  10.1103/PhysRevB.92.060413} {\bibfield  {journal} {\bibinfo  {journal}
  {Physical Review B}\ }\textbf {\bibinfo {volume} {92}},\ \bibinfo {pages}
  {060413} (\bibinfo {year} {2015})}\BibitemShut {NoStop}%
\bibitem [{\citenamefont {{Sl{\"o}etjes}}\ \emph {et~al.}(2021)\citenamefont
  {{Sl{\"o}etjes}}, \citenamefont {{Hj{\"o}rvarsson}},\ and\ \citenamefont
  {{Kapaklis}}}]{Sloetjes_arXiv_2020}%
  \BibitemOpen
  \bibfield  {author} {\bibinfo {author} {\bibfnamefont {S.~D.}\ \bibnamefont
  {{Sl{\"o}etjes}}}, \bibinfo {author} {\bibfnamefont {B.}~\bibnamefont
  {{Hj{\"o}rvarsson}}}, \ and\ \bibinfo {author} {\bibfnamefont
  {V.}~\bibnamefont {{Kapaklis}}},\ }\href {\doibase 10.1063/5.0048789}
  {\bibfield  {journal} {\bibinfo  {journal} {Applied Physics Letters}\
  }\textbf {\bibinfo {volume} {118}},\ \bibinfo {pages} {142407} (\bibinfo
  {year} {2021})}\BibitemShut {NoStop}%
\bibitem [{\citenamefont {Sendetskyi}\ \emph {et~al.}(2019)\citenamefont
  {Sendetskyi}, \citenamefont {Scagnoli}, \citenamefont {Leo}, \citenamefont
  {Anghinolfi}, \citenamefont {Alberca}, \citenamefont {L{\"u}ning},
  \citenamefont {Staub}, \citenamefont {Derlet},\ and\ \citenamefont
  {Heyderman}}]{sendetskyi_continuous_2019}%
  \BibitemOpen
  \bibfield  {author} {\bibinfo {author} {\bibfnamefont {O.}~\bibnamefont
  {Sendetskyi}}, \bibinfo {author} {\bibfnamefont {V.}~\bibnamefont
  {Scagnoli}}, \bibinfo {author} {\bibfnamefont {N.}~\bibnamefont {Leo}},
  \bibinfo {author} {\bibfnamefont {L.}~\bibnamefont {Anghinolfi}}, \bibinfo
  {author} {\bibfnamefont {A.}~\bibnamefont {Alberca}}, \bibinfo {author}
  {\bibfnamefont {J.}~\bibnamefont {L{\"u}ning}}, \bibinfo {author}
  {\bibfnamefont {U.}~\bibnamefont {Staub}}, \bibinfo {author} {\bibfnamefont
  {P.~M.}\ \bibnamefont {Derlet}}, \ and\ \bibinfo {author} {\bibfnamefont
  {L.~J.}\ \bibnamefont {Heyderman}},\ }\href {\doibase
  10.1103/PhysRevB.99.214430} {\bibfield  {journal} {\bibinfo  {journal}
  {Physical Review B}\ }\textbf {\bibinfo {volume} {99}},\ \bibinfo {pages}
  {214430} (\bibinfo {year} {2019})}\BibitemShut {NoStop}%
\bibitem [{\citenamefont {Ewerlin}\ \emph {et~al.}(2013)\citenamefont
  {Ewerlin}, \citenamefont {Demirbas}, \citenamefont {Br{\"u}ssing},
  \citenamefont {Petracic}, \citenamefont {{\"U}nal}, \citenamefont {Valencia},
  \citenamefont {Kronast},\ and\ \citenamefont
  {Zabel}}]{ewerlin_magnetic_2013}%
  \BibitemOpen
  \bibfield  {author} {\bibinfo {author} {\bibfnamefont {M.}~\bibnamefont
  {Ewerlin}}, \bibinfo {author} {\bibfnamefont {D.}~\bibnamefont {Demirbas}},
  \bibinfo {author} {\bibfnamefont {F.}~\bibnamefont {Br{\"u}ssing}}, \bibinfo
  {author} {\bibfnamefont {O.}~\bibnamefont {Petracic}}, \bibinfo {author}
  {\bibfnamefont {A.~A.}\ \bibnamefont {{\"U}nal}}, \bibinfo {author}
  {\bibfnamefont {S.}~\bibnamefont {Valencia}}, \bibinfo {author}
  {\bibfnamefont {F.}~\bibnamefont {Kronast}}, \ and\ \bibinfo {author}
  {\bibfnamefont {H.}~\bibnamefont {Zabel}},\ }\href {\doibase
  10.1103/PhysRevLett.110.177209} {\bibfield  {journal} {\bibinfo  {journal}
  {Physical Review Letters}\ }\textbf {\bibinfo {volume} {110}},\ \bibinfo
  {pages} {177209} (\bibinfo {year} {2013})}\BibitemShut {NoStop}%
\bibitem [{\citenamefont {Streubel}\ \emph {et~al.}(2018)\citenamefont
  {Streubel}, \citenamefont {Kent}, \citenamefont {Dhuey}, \citenamefont
  {Scholl}, \citenamefont {Kevan},\ and\ \citenamefont
  {Fischer}}]{streubel_spatial_2018}%
  \BibitemOpen
  \bibfield  {author} {\bibinfo {author} {\bibfnamefont {R.}~\bibnamefont
  {Streubel}}, \bibinfo {author} {\bibfnamefont {N.}~\bibnamefont {Kent}},
  \bibinfo {author} {\bibfnamefont {S.}~\bibnamefont {Dhuey}}, \bibinfo
  {author} {\bibfnamefont {A.}~\bibnamefont {Scholl}}, \bibinfo {author}
  {\bibfnamefont {S.}~\bibnamefont {Kevan}}, \ and\ \bibinfo {author}
  {\bibfnamefont {P.}~\bibnamefont {Fischer}},\ }\href {\doibase
  10.1021/acs.nanolett.8b01789} {\bibfield  {journal} {\bibinfo  {journal}
  {Nano Letters}\ }\textbf {\bibinfo {volume} {18}},\ \bibinfo {pages} {7428}
  (\bibinfo {year} {2018})}\BibitemShut {NoStop}%
\bibitem [{\citenamefont {Leo}\ \emph {et~al.}(2018)\citenamefont {Leo},
  \citenamefont {Holenstein}, \citenamefont {Schildknecht}, \citenamefont
  {Sendetskyi}, \citenamefont {Luetkens}, \citenamefont {Derlet}, \citenamefont
  {Scagnoli}, \citenamefont {Lan{\c c}on}, \citenamefont {Mardegan},
  \citenamefont {Prokscha}, \citenamefont {Suter}, \citenamefont {Salman},
  \citenamefont {Lee},\ and\ \citenamefont {Heyderman}}]{leo_collective_2018}%
  \BibitemOpen
  \bibfield  {author} {\bibinfo {author} {\bibfnamefont {N.}~\bibnamefont
  {Leo}}, \bibinfo {author} {\bibfnamefont {S.}~\bibnamefont {Holenstein}},
  \bibinfo {author} {\bibfnamefont {D.}~\bibnamefont {Schildknecht}}, \bibinfo
  {author} {\bibfnamefont {O.}~\bibnamefont {Sendetskyi}}, \bibinfo {author}
  {\bibfnamefont {H.}~\bibnamefont {Luetkens}}, \bibinfo {author}
  {\bibfnamefont {P.~M.}\ \bibnamefont {Derlet}}, \bibinfo {author}
  {\bibfnamefont {V.}~\bibnamefont {Scagnoli}}, \bibinfo {author}
  {\bibfnamefont {D.}~\bibnamefont {Lan{\c c}on}}, \bibinfo {author}
  {\bibfnamefont {J.~R.~L.}\ \bibnamefont {Mardegan}}, \bibinfo {author}
  {\bibfnamefont {T.}~\bibnamefont {Prokscha}}, \bibinfo {author}
  {\bibfnamefont {A.}~\bibnamefont {Suter}}, \bibinfo {author} {\bibfnamefont
  {Z.}~\bibnamefont {Salman}}, \bibinfo {author} {\bibfnamefont
  {S.}~\bibnamefont {Lee}}, \ and\ \bibinfo {author} {\bibfnamefont {L.~J.}\
  \bibnamefont {Heyderman}},\ }\href {\doibase 10.1038/s41467-018-05216-2}
  {\bibfield  {journal} {\bibinfo  {journal} {Nature Communications}\ }\textbf
  {\bibinfo {volume} {9}},\ \bibinfo {pages} {2850} (\bibinfo {year}
  {2018})}\BibitemShut {NoStop}%
\bibitem [{\citenamefont {Arnalds}\ \emph {et~al.}(2016)\citenamefont
  {Arnalds}, \citenamefont {Chico}, \citenamefont {Stopfel}, \citenamefont
  {Kapaklis}, \citenamefont {B{\"a}renbold}, \citenamefont {Verschuuren},
  \citenamefont {Wolff}, \citenamefont {Neu}, \citenamefont {Bergman},\ and\
  \citenamefont {Hj{\"o}rvarsson}}]{arnalds2016new}%
  \BibitemOpen
  \bibfield  {author} {\bibinfo {author} {\bibfnamefont {U.~B.}\ \bibnamefont
  {Arnalds}}, \bibinfo {author} {\bibfnamefont {J.}~\bibnamefont {Chico}},
  \bibinfo {author} {\bibfnamefont {H.}~\bibnamefont {Stopfel}}, \bibinfo
  {author} {\bibfnamefont {V.}~\bibnamefont {Kapaklis}}, \bibinfo {author}
  {\bibfnamefont {O.}~\bibnamefont {B{\"a}renbold}}, \bibinfo {author}
  {\bibfnamefont {M.~A.}\ \bibnamefont {Verschuuren}}, \bibinfo {author}
  {\bibfnamefont {U.}~\bibnamefont {Wolff}}, \bibinfo {author} {\bibfnamefont
  {V.}~\bibnamefont {Neu}}, \bibinfo {author} {\bibfnamefont {A.}~\bibnamefont
  {Bergman}}, \ and\ \bibinfo {author} {\bibfnamefont {B.}~\bibnamefont
  {Hj{\"o}rvarsson}},\ }\href@noop {} {\bibfield  {journal} {\bibinfo
  {journal} {New Journal of Physics}\ }\textbf {\bibinfo {volume} {18}},\
  \bibinfo {pages} {023008} (\bibinfo {year} {2016})}\BibitemShut {NoStop}%
\bibitem [{\citenamefont {Arnalds}\ \emph {et~al.}(2014)\citenamefont
  {Arnalds}, \citenamefont {Ahlberg}, \citenamefont {Brewer}, \citenamefont
  {Kapaklis}, \citenamefont {Papaioannou}, \citenamefont {Karimipour},
  \citenamefont {Korelis}, \citenamefont {Stein}, \citenamefont {{\'O}lafsson},
  \citenamefont {Hase},\ and\ \citenamefont {Hj{\"o}rvarsson}}]{Arnalds_XY}%
  \BibitemOpen
  \bibfield  {author} {\bibinfo {author} {\bibfnamefont {U.~B.}\ \bibnamefont
  {Arnalds}}, \bibinfo {author} {\bibfnamefont {M.}~\bibnamefont {Ahlberg}},
  \bibinfo {author} {\bibfnamefont {M.~S.}\ \bibnamefont {Brewer}}, \bibinfo
  {author} {\bibfnamefont {V.}~\bibnamefont {Kapaklis}}, \bibinfo {author}
  {\bibfnamefont {E.~T.}\ \bibnamefont {Papaioannou}}, \bibinfo {author}
  {\bibfnamefont {M.}~\bibnamefont {Karimipour}}, \bibinfo {author}
  {\bibfnamefont {P.}~\bibnamefont {Korelis}}, \bibinfo {author} {\bibfnamefont
  {A.}~\bibnamefont {Stein}}, \bibinfo {author} {\bibfnamefont
  {S.}~\bibnamefont {{\'O}lafsson}}, \bibinfo {author} {\bibfnamefont
  {T.~P.~A.}\ \bibnamefont {Hase}}, \ and\ \bibinfo {author} {\bibfnamefont
  {B.}~\bibnamefont {Hj{\"o}rvarsson}},\ }\href {\doibase 10.1063/1.4891479}
  {\bibfield  {journal} {\bibinfo  {journal} {Applied Physics Letters}\
  }\textbf {\bibinfo {volume} {105}},\ \bibinfo {pages} {042409} (\bibinfo
  {year} {2014})}\BibitemShut {NoStop}%
\bibitem [{\citenamefont {{\"O}stman}\ \emph
  {et~al.}(2018{\natexlab{b}})\citenamefont {{\"O}stman}, \citenamefont
  {Stopfel}, \citenamefont {Chioar}, \citenamefont {Arnalds}, \citenamefont
  {Stein}, \citenamefont {Kapaklis},\ and\ \citenamefont
  {Hj{\"o}rvarsson}}]{ostman_interaction_2018}%
  \BibitemOpen
  \bibfield  {author} {\bibinfo {author} {\bibfnamefont {E.}~\bibnamefont
  {{\"O}stman}}, \bibinfo {author} {\bibfnamefont {H.}~\bibnamefont {Stopfel}},
  \bibinfo {author} {\bibfnamefont {I.-A.}\ \bibnamefont {Chioar}}, \bibinfo
  {author} {\bibfnamefont {U.~B.}\ \bibnamefont {Arnalds}}, \bibinfo {author}
  {\bibfnamefont {A.}~\bibnamefont {Stein}}, \bibinfo {author} {\bibfnamefont
  {V.}~\bibnamefont {Kapaklis}}, \ and\ \bibinfo {author} {\bibfnamefont
  {B.}~\bibnamefont {Hj{\"o}rvarsson}},\ }\href {\doibase
  10.1038/s41567-017-0027-2} {\bibfield  {journal} {\bibinfo  {journal} {Nature
  Physics}\ }\textbf {\bibinfo {volume} {14}},\ \bibinfo {pages} {375}
  (\bibinfo {year} {2018}{\natexlab{b}})}\BibitemShut {NoStop}%
\bibitem [{\citenamefont {Shinjo}(2000)}]{shinjo_magnetic_2000}%
  \BibitemOpen
  \bibfield  {author} {\bibinfo {author} {\bibfnamefont {T.}~\bibnamefont
  {Shinjo}},\ }\href {\doibase 10.1126/science.289.5481.930} {\bibfield
  {journal} {\bibinfo  {journal} {Science}\ }\textbf {\bibinfo {volume}
  {289}},\ \bibinfo {pages} {930} (\bibinfo {year} {2000})}\BibitemShut
  {NoStop}%
\bibitem [{\citenamefont {Kl{\"a}ui}\ \emph {et~al.}(2003)\citenamefont
  {Kl{\"a}ui}, \citenamefont {Vaz}, \citenamefont {Lopez-Diaz},\ and\
  \citenamefont {Bland}}]{Klaui_vortx_2003}%
  \BibitemOpen
  \bibfield  {author} {\bibinfo {author} {\bibfnamefont {M.}~\bibnamefont
  {Kl{\"a}ui}}, \bibinfo {author} {\bibfnamefont {C.~A.~F.}\ \bibnamefont
  {Vaz}}, \bibinfo {author} {\bibfnamefont {L.}~\bibnamefont {Lopez-Diaz}}, \
  and\ \bibinfo {author} {\bibfnamefont {J.~A.~C.}\ \bibnamefont {Bland}},\
  }\href {\doibase 10.1088/0953-8984/15/21/201} {\bibfield  {journal} {\bibinfo
   {journal} {Journal of Physics: Condensed Matter}\ }\textbf {\bibinfo
  {volume} {15}},\ \bibinfo {pages} {R985} (\bibinfo {year}
  {2003})}\BibitemShut {NoStop}%
\bibitem [{\citenamefont {Mermin}(1979)}]{Mermin:1979io}%
  \BibitemOpen
  \bibfield  {author} {\bibinfo {author} {\bibfnamefont {N.~D.}\ \bibnamefont
  {Mermin}},\ }\href {\doibase 10.1103/RevModPhys.51.591} {\bibfield  {journal}
  {\bibinfo  {journal} {Reviews of Modern Physics}\ }\textbf {\bibinfo {volume}
  {51}},\ \bibinfo {pages} {591} (\bibinfo {year} {1979})}\BibitemShut
  {NoStop}%
\bibitem [{\citenamefont {Tchernyshyov}\ and\ \citenamefont
  {Chern}(2005)}]{Tchernyshyov:2005gs}%
  \BibitemOpen
  \bibfield  {author} {\bibinfo {author} {\bibfnamefont {O.}~\bibnamefont
  {Tchernyshyov}}\ and\ \bibinfo {author} {\bibfnamefont {G.-W.}\ \bibnamefont
  {Chern}},\ }\href {\doibase 10.1103/PhysRevLett.95.197204} {\bibfield
  {journal} {\bibinfo  {journal} {Physical Review Letters}\ }\textbf {\bibinfo
  {volume} {95}},\ \bibinfo {pages} {197204} (\bibinfo {year}
  {2005})}\BibitemShut {NoStop}%
\bibitem [{\citenamefont {{\"O}stman}\ \emph {et~al.}(2014)\citenamefont
  {{\"O}stman}, \citenamefont {Arnalds}, \citenamefont {Melander},
  \citenamefont {Kapaklis}, \citenamefont {P{\'a}lsson}, \citenamefont {Saw},
  \citenamefont {Verschuuren}, \citenamefont {Kronast}, \citenamefont
  {Papaioannou}, \citenamefont {Fadley},\ and\ \citenamefont
  {Hj{\"o}rvarsson}}]{ostman_hysteresis-free_2014}%
  \BibitemOpen
  \bibfield  {author} {\bibinfo {author} {\bibfnamefont {E.}~\bibnamefont
  {{\"O}stman}}, \bibinfo {author} {\bibfnamefont {U.~B.}\ \bibnamefont
  {Arnalds}}, \bibinfo {author} {\bibfnamefont {E.}~\bibnamefont {Melander}},
  \bibinfo {author} {\bibfnamefont {V.}~\bibnamefont {Kapaklis}}, \bibinfo
  {author} {\bibfnamefont {G.~K.}\ \bibnamefont {P{\'a}lsson}}, \bibinfo
  {author} {\bibfnamefont {A.~Y.}\ \bibnamefont {Saw}}, \bibinfo {author}
  {\bibfnamefont {M.~A.}\ \bibnamefont {Verschuuren}}, \bibinfo {author}
  {\bibfnamefont {F.}~\bibnamefont {Kronast}}, \bibinfo {author} {\bibfnamefont
  {E.~T.}\ \bibnamefont {Papaioannou}}, \bibinfo {author} {\bibfnamefont
  {C.~S.}\ \bibnamefont {Fadley}}, \ and\ \bibinfo {author} {\bibfnamefont
  {B.}~\bibnamefont {Hj{\"o}rvarsson}},\ }\href {\doibase
  10.1088/1367-2630/16/5/053002} {\bibfield  {journal} {\bibinfo  {journal}
  {New Journal of Physics}\ }\textbf {\bibinfo {volume} {16}},\ \bibinfo
  {pages} {053002} (\bibinfo {year} {2014})}\BibitemShut {NoStop}%
\bibitem [{\citenamefont {Guslienko}\ \emph {et~al.}(2001)\citenamefont
  {Guslienko}, \citenamefont {Novosad}, \citenamefont {Otani}, \citenamefont
  {Shima},\ and\ \citenamefont {Fukamichi}}]{guslienko_coupling_2001}%
  \BibitemOpen
  \bibfield  {author} {\bibinfo {author} {\bibfnamefont {K.~Y.}\ \bibnamefont
  {Guslienko}}, \bibinfo {author} {\bibfnamefont {V.}~\bibnamefont {Novosad}},
  \bibinfo {author} {\bibfnamefont {Y.}~\bibnamefont {Otani}}, \bibinfo
  {author} {\bibfnamefont {H.}~\bibnamefont {Shima}}, \ and\ \bibinfo {author}
  {\bibfnamefont {K.}~\bibnamefont {Fukamichi}},\ }\href {\doibase
  10.1103/PhysRevB.65.024414} {\bibfield  {journal} {\bibinfo  {journal}
  {Physical Review B}\ }\textbf {\bibinfo {volume} {65}},\ \bibinfo {pages}
  {024414} (\bibinfo {year} {2001})}\BibitemShut {NoStop}%
\bibitem [{\citenamefont {Belkhou}\ \emph {et~al.}(2015)\citenamefont
  {Belkhou}, \citenamefont {Stanescu}, \citenamefont {Swaraj}, \citenamefont
  {Besson}, \citenamefont {Ledoux}, \citenamefont {Hajlaoui},\ and\
  \citenamefont {Dalle}}]{belkhou_hermes_2015}%
  \BibitemOpen
  \bibfield  {author} {\bibinfo {author} {\bibfnamefont {R.}~\bibnamefont
  {Belkhou}}, \bibinfo {author} {\bibfnamefont {S.}~\bibnamefont {Stanescu}},
  \bibinfo {author} {\bibfnamefont {S.}~\bibnamefont {Swaraj}}, \bibinfo
  {author} {\bibfnamefont {A.}~\bibnamefont {Besson}}, \bibinfo {author}
  {\bibfnamefont {M.}~\bibnamefont {Ledoux}}, \bibinfo {author} {\bibfnamefont
  {M.}~\bibnamefont {Hajlaoui}}, \ and\ \bibinfo {author} {\bibfnamefont
  {D.}~\bibnamefont {Dalle}},\ }\href {\doibase 10.1107/S1600577515007778}
  {\bibfield  {journal} {\bibinfo  {journal} {Journal of Synchrotron
  Radiation}\ }\textbf {\bibinfo {volume} {22}},\ \bibinfo {pages} {968}
  (\bibinfo {year} {2015})}\BibitemShut {NoStop}%
\bibitem [{\citenamefont {Doran}\ \emph {et~al.}(2012)\citenamefont {Doran},
  \citenamefont {Church}, \citenamefont {Miller}, \citenamefont {Morrison},
  \citenamefont {Young},\ and\ \citenamefont {Scholl}}]{doran_cryogenic_2012}%
  \BibitemOpen
  \bibfield  {author} {\bibinfo {author} {\bibfnamefont {A.}~\bibnamefont
  {Doran}}, \bibinfo {author} {\bibfnamefont {M.}~\bibnamefont {Church}},
  \bibinfo {author} {\bibfnamefont {T.}~\bibnamefont {Miller}}, \bibinfo
  {author} {\bibfnamefont {G.}~\bibnamefont {Morrison}}, \bibinfo {author}
  {\bibfnamefont {A.~T.}\ \bibnamefont {Young}}, \ and\ \bibinfo {author}
  {\bibfnamefont {A.}~\bibnamefont {Scholl}},\ }\href {\doibase
  10.1016/j.elspec.2012.05.005} {\bibfield  {journal} {\bibinfo  {journal}
  {Journal of Electron Spectroscopy and Related Phenomena}\ }\textbf {\bibinfo
  {volume} {185}},\ \bibinfo {pages} {340} (\bibinfo {year}
  {2012})}\BibitemShut {NoStop}%
\bibitem [{\citenamefont {Vansteenkiste}\ \emph {et~al.}(2014)\citenamefont
  {Vansteenkiste}, \citenamefont {Leliaert}, \citenamefont {Dvornik},
  \citenamefont {Helsen}, \citenamefont {Garcia-Sanchez},\ and\ \citenamefont
  {Van~Waeyenberge}}]{vansteenkiste_design_2014}%
  \BibitemOpen
  \bibfield  {author} {\bibinfo {author} {\bibfnamefont {A.}~\bibnamefont
  {Vansteenkiste}}, \bibinfo {author} {\bibfnamefont {J.}~\bibnamefont
  {Leliaert}}, \bibinfo {author} {\bibfnamefont {M.}~\bibnamefont {Dvornik}},
  \bibinfo {author} {\bibfnamefont {M.}~\bibnamefont {Helsen}}, \bibinfo
  {author} {\bibfnamefont {F.}~\bibnamefont {Garcia-Sanchez}}, \ and\ \bibinfo
  {author} {\bibfnamefont {B.}~\bibnamefont {Van~Waeyenberge}},\ }\href
  {\doibase 10.1063/1.4899186} {\bibfield  {journal} {\bibinfo  {journal} {AIP
  Advances}\ }\textbf {\bibinfo {volume} {4}},\ \bibinfo {pages} {107133}
  (\bibinfo {year} {2014})}\BibitemShut {NoStop}%
\bibitem [{\citenamefont {Ciuciulkaite}\ \emph {et~al.}(2019)\citenamefont
  {Ciuciulkaite}, \citenamefont {{\"O}stman}, \citenamefont {Brucas},
  \citenamefont {Kumar}, \citenamefont {Verschuuren}, \citenamefont
  {Svedlindh}, \citenamefont {Hj{\"o}rvarsson},\ and\ \citenamefont
  {Kapaklis}}]{ciuciulkaite_collective_2019}%
  \BibitemOpen
  \bibfield  {author} {\bibinfo {author} {\bibfnamefont {A.}~\bibnamefont
  {Ciuciulkaite}}, \bibinfo {author} {\bibfnamefont {E.}~\bibnamefont
  {{\"O}stman}}, \bibinfo {author} {\bibfnamefont {R.}~\bibnamefont {Brucas}},
  \bibinfo {author} {\bibfnamefont {A.}~\bibnamefont {Kumar}}, \bibinfo
  {author} {\bibfnamefont {M.~A.}\ \bibnamefont {Verschuuren}}, \bibinfo
  {author} {\bibfnamefont {P.}~\bibnamefont {Svedlindh}}, \bibinfo {author}
  {\bibfnamefont {B.}~\bibnamefont {Hj{\"o}rvarsson}}, \ and\ \bibinfo {author}
  {\bibfnamefont {V.}~\bibnamefont {Kapaklis}},\ }\href {\doibase
  10.1103/PhysRevB.99.184415} {\bibfield  {journal} {\bibinfo  {journal}
  {Physical Review B}\ }\textbf {\bibinfo {volume} {99}},\ \bibinfo {pages}
  {184415} (\bibinfo {year} {2019})}\BibitemShut {NoStop}%
\bibitem [{\citenamefont {Ding}\ \emph {et~al.}(2005)\citenamefont {Ding},
  \citenamefont {Schmid}, \citenamefont {Li}, \citenamefont {Guslienko},\ and\
  \citenamefont {Bader}}]{ding_magnetic_2005}%
  \BibitemOpen
  \bibfield  {author} {\bibinfo {author} {\bibfnamefont {H.~F.}\ \bibnamefont
  {Ding}}, \bibinfo {author} {\bibfnamefont {A.~K.}\ \bibnamefont {Schmid}},
  \bibinfo {author} {\bibfnamefont {D.}~\bibnamefont {Li}}, \bibinfo {author}
  {\bibfnamefont {K.~Y.}\ \bibnamefont {Guslienko}}, \ and\ \bibinfo {author}
  {\bibfnamefont {S.~D.}\ \bibnamefont {Bader}},\ }\href {\doibase
  10.1103/PhysRevLett.94.157202} {\bibfield  {journal} {\bibinfo  {journal}
  {Physical Review Letters}\ }\textbf {\bibinfo {volume} {94}},\ \bibinfo
  {pages} {157202} (\bibinfo {year} {2005})}\BibitemShut {NoStop}%
\bibitem [{\citenamefont {Zhang}\ \emph {et~al.}(2015)\citenamefont {Zhang},
  \citenamefont {Baker}, \citenamefont {Komineas},\ and\ \citenamefont
  {Hesjedal}}]{Zhang:2015de}%
  \BibitemOpen
  \bibfield  {author} {\bibinfo {author} {\bibfnamefont {S.}~\bibnamefont
  {Zhang}}, \bibinfo {author} {\bibfnamefont {A.~A.}\ \bibnamefont {Baker}},
  \bibinfo {author} {\bibfnamefont {S.}~\bibnamefont {Komineas}}, \ and\
  \bibinfo {author} {\bibfnamefont {T.}~\bibnamefont {Hesjedal}},\ }\href
  {\doibase 10.1038/srep15773} {\bibfield  {journal} {\bibinfo  {journal}
  {Scientific Reports}\ }\textbf {\bibinfo {volume} {5}},\ \bibinfo {pages}
  {591} (\bibinfo {year} {2015})}\BibitemShut {NoStop}%
\bibitem [{\citenamefont {Donnelly}\ \emph {et~al.}(2020)\citenamefont
  {Donnelly}, \citenamefont {Metlov}, \citenamefont {Scagnoli}, \citenamefont
  {Guizar-Sicairos}, \citenamefont {Holler}, \citenamefont {Bingham},
  \citenamefont {Raabe}, \citenamefont {Heyderman}, \citenamefont {Cooper},\
  and\ \citenamefont {Gliga}}]{Donnelly_2020fh}%
  \BibitemOpen
  \bibfield  {author} {\bibinfo {author} {\bibfnamefont {C.}~\bibnamefont
  {Donnelly}}, \bibinfo {author} {\bibfnamefont {K.~L.}\ \bibnamefont
  {Metlov}}, \bibinfo {author} {\bibfnamefont {V.}~\bibnamefont {Scagnoli}},
  \bibinfo {author} {\bibfnamefont {M.}~\bibnamefont {Guizar-Sicairos}},
  \bibinfo {author} {\bibfnamefont {M.}~\bibnamefont {Holler}}, \bibinfo
  {author} {\bibfnamefont {N.~S.}\ \bibnamefont {Bingham}}, \bibinfo {author}
  {\bibfnamefont {J.}~\bibnamefont {Raabe}}, \bibinfo {author} {\bibfnamefont
  {L.~J.}\ \bibnamefont {Heyderman}}, \bibinfo {author} {\bibfnamefont {N.~R.}\
  \bibnamefont {Cooper}}, \ and\ \bibinfo {author} {\bibfnamefont
  {S.}~\bibnamefont {Gliga}},\ }\href {\doibase 10.1038/s41567-020-01057-3}
  {\bibfield  {journal} {\bibinfo  {journal} {Nature Physics}\ }\textbf
  {\bibinfo {volume} {21}},\ \bibinfo {pages} {759{\textendash}761} (\bibinfo
  {year} {2020})}\BibitemShut {NoStop}%
\bibitem [{\citenamefont {Sl{\"o}etjes}\ \emph {et~al.}(2020)\citenamefont
  {Sl{\"o}etjes}, \citenamefont {Folven},\ and\ \citenamefont
  {Grepstad}}]{Sloetjes:2020iw}%
  \BibitemOpen
  \bibfield  {author} {\bibinfo {author} {\bibfnamefont {S.~D.}\ \bibnamefont
  {Sl{\"o}etjes}}, \bibinfo {author} {\bibfnamefont {E.}~\bibnamefont
  {Folven}}, \ and\ \bibinfo {author} {\bibfnamefont {J.~K.}\ \bibnamefont
  {Grepstad}},\ }\href {\doibase 10.1103/PhysRevB.101.014450} {\bibfield
  {journal} {\bibinfo  {journal} {Physical Review B}\ }\textbf {\bibinfo
  {volume} {101}},\ \bibinfo {pages} {014450} (\bibinfo {year}
  {2020})}\BibitemShut {NoStop}%
\bibitem [{\citenamefont {Cowburn}\ \emph {et~al.}(1999)\citenamefont
  {Cowburn}, \citenamefont {Koltsov}, \citenamefont {Adeyeye}, \citenamefont
  {Welland},\ and\ \citenamefont {Tricker}}]{cowburn_single-domain_1999}%
  \BibitemOpen
  \bibfield  {author} {\bibinfo {author} {\bibfnamefont {R.~P.}\ \bibnamefont
  {Cowburn}}, \bibinfo {author} {\bibfnamefont {D.~K.}\ \bibnamefont
  {Koltsov}}, \bibinfo {author} {\bibfnamefont {A.~O.}\ \bibnamefont
  {Adeyeye}}, \bibinfo {author} {\bibfnamefont {M.~E.}\ \bibnamefont
  {Welland}}, \ and\ \bibinfo {author} {\bibfnamefont {D.~M.}\ \bibnamefont
  {Tricker}},\ }\href {\doibase 10.1103/PhysRevLett.83.1042} {\bibfield
  {journal} {\bibinfo  {journal} {Physical Review Letters}\ }\textbf {\bibinfo
  {volume} {83}},\ \bibinfo {pages} {1042} (\bibinfo {year}
  {1999})}\BibitemShut {NoStop}%
\bibitem [{\citenamefont {Wilson}(1979)}]{Wilson:1979wn}%
  \BibitemOpen
  \bibfield  {author} {\bibinfo {author} {\bibfnamefont {K.~G.}\ \bibnamefont
  {Wilson}},\ }\href {\doibase 10.1038/scientificamerican0879-158} {\bibfield
  {journal} {\bibinfo  {journal} {Scientific American}\ }\textbf {\bibinfo
  {volume} {241}},\ \bibinfo {pages} {158} (\bibinfo {year}
  {1979})}\BibitemShut {NoStop}%
\bibitem [{\citenamefont {Castellano}\ \emph {et~al.}(2009)\citenamefont
  {Castellano}, \citenamefont {Fortunato},\ and\ \citenamefont
  {Loreto}}]{castellano2009statistical}%
  \BibitemOpen
  \bibfield  {author} {\bibinfo {author} {\bibfnamefont {C.}~\bibnamefont
  {Castellano}}, \bibinfo {author} {\bibfnamefont {S.}~\bibnamefont
  {Fortunato}}, \ and\ \bibinfo {author} {\bibfnamefont {V.}~\bibnamefont
  {Loreto}},\ }\href@noop {} {\bibfield  {journal} {\bibinfo  {journal}
  {Reviews of modern physics}\ }\textbf {\bibinfo {volume} {81}},\ \bibinfo
  {pages} {591} (\bibinfo {year} {2009})}\BibitemShut {NoStop}%
\bibitem [{\citenamefont {Michaud}\ and\ \citenamefont
  {Szilva}(2018)}]{michaud2018social}%
  \BibitemOpen
  \bibfield  {author} {\bibinfo {author} {\bibfnamefont {J.}~\bibnamefont
  {Michaud}}\ and\ \bibinfo {author} {\bibfnamefont {A.}~\bibnamefont
  {Szilva}},\ }\href@noop {} {\bibfield  {journal} {\bibinfo  {journal}
  {Physical Review E}\ }\textbf {\bibinfo {volume} {97}},\ \bibinfo {pages}
  {062313} (\bibinfo {year} {2018})}\BibitemShut {NoStop}%
\bibitem [{\citenamefont {Jenkins}\ \emph {et~al.}(2014)\citenamefont
  {Jenkins}, \citenamefont {Grimaldi}, \citenamefont {Bortolotti},
  \citenamefont {Lebrun}, \citenamefont {Kubota}, \citenamefont {Yakushiji},
  \citenamefont {Fukushima}, \citenamefont {de~Loubens}, \citenamefont {Klein},
  \citenamefont {Yuasa},\ and\ \citenamefont {Cros}}]{Jenkins_vortices_2014}%
  \BibitemOpen
  \bibfield  {author} {\bibinfo {author} {\bibfnamefont {A.~S.}\ \bibnamefont
  {Jenkins}}, \bibinfo {author} {\bibfnamefont {E.}~\bibnamefont {Grimaldi}},
  \bibinfo {author} {\bibfnamefont {P.}~\bibnamefont {Bortolotti}}, \bibinfo
  {author} {\bibfnamefont {R.}~\bibnamefont {Lebrun}}, \bibinfo {author}
  {\bibfnamefont {H.}~\bibnamefont {Kubota}}, \bibinfo {author} {\bibfnamefont
  {K.}~\bibnamefont {Yakushiji}}, \bibinfo {author} {\bibfnamefont
  {A.}~\bibnamefont {Fukushima}}, \bibinfo {author} {\bibfnamefont
  {G.}~\bibnamefont {de~Loubens}}, \bibinfo {author} {\bibfnamefont
  {O.}~\bibnamefont {Klein}}, \bibinfo {author} {\bibfnamefont
  {S.}~\bibnamefont {Yuasa}}, \ and\ \bibinfo {author} {\bibfnamefont
  {V.}~\bibnamefont {Cros}},\ }\href {\doibase 10.1063/1.4900743} {\bibfield
  {journal} {\bibinfo  {journal} {Applied Physics Letters}\ }\textbf {\bibinfo
  {volume} {105}},\ \bibinfo {pages} {172403} (\bibinfo {year}
  {2014})}\BibitemShut {NoStop}%
\bibitem [{\citenamefont {Pohlit}\ \emph {et~al.}(2020)\citenamefont {Pohlit},
  \citenamefont {Muscas}, \citenamefont {Chioar}, \citenamefont {Stopfel},
  \citenamefont {Ciuciulkaite}, \citenamefont {{\"O}stman}, \citenamefont
  {Pappas}, \citenamefont {Stein}, \citenamefont {Hj{\"o}rvarsson},
  \citenamefont {J{\"o}nsson},\ and\ \citenamefont
  {Kapaklis}}]{Pohlit_PRB_2020}%
  \BibitemOpen
  \bibfield  {author} {\bibinfo {author} {\bibfnamefont {M.}~\bibnamefont
  {Pohlit}}, \bibinfo {author} {\bibfnamefont {G.}~\bibnamefont {Muscas}},
  \bibinfo {author} {\bibfnamefont {I.-A.}\ \bibnamefont {Chioar}}, \bibinfo
  {author} {\bibfnamefont {H.}~\bibnamefont {Stopfel}}, \bibinfo {author}
  {\bibfnamefont {A.}~\bibnamefont {Ciuciulkaite}}, \bibinfo {author}
  {\bibfnamefont {E.}~\bibnamefont {{\"O}stman}}, \bibinfo {author}
  {\bibfnamefont {S.~D.}\ \bibnamefont {Pappas}}, \bibinfo {author}
  {\bibfnamefont {A.}~\bibnamefont {Stein}}, \bibinfo {author} {\bibfnamefont
  {B.}~\bibnamefont {Hj{\"o}rvarsson}}, \bibinfo {author} {\bibfnamefont
  {P.~E.}\ \bibnamefont {J{\"o}nsson}}, \ and\ \bibinfo {author} {\bibfnamefont
  {V.}~\bibnamefont {Kapaklis}},\ }\href {\doibase 10.1103/PhysRevB.101.134404}
  {\bibfield  {journal} {\bibinfo  {journal} {Physical Review B}\ }\textbf
  {\bibinfo {volume} {101}},\ \bibinfo {pages} {134404} (\bibinfo {year}
  {2020})}\BibitemShut {NoStop}%
\end{thebibliography}

%

\end{document}